\documentclass[smallextended]{svjour3}       % onecolumn (second format)
\smartqed  % flush right qed marks, e.g. at end of proof
\usepackage{graphicx}
\usepackage{float}
\usepackage{amsfonts}
\usepackage{amsmath}
\usepackage{subcaption}
\begin{document}

\title{Does Subjective Well-being Contribute to Our Understanding of Mexican Well-being?
}

\titlerunning{Subjective Well-being}        % if too long for running head

\author{Jeremy Heald
	\and Erick Trevi\~no Aguilar. 
}


\institute{J.H. \at
              	Departament of Economics and Finance, University of Guanajuato, M\'exico. Email: healdj59@gmail.com
           \and
           E.T.A. \at
              Institute for Mathematics, UNAM. 
              \email{ erick.trevino@im.unam.mx}
}

\date{Received: date / Accepted: date}

\maketitle

\begin{abstract}
The article reviews the history of well-being to gauge how subjective question surveys can improve our understanding of well-being in Mexico. The research uses data at the level of the 32 federal entities or States, taking advantage of the heterogeneity in development indicator readings between and within geographical areas, the product of socioeconomic inequality. The data come principally from two innovative subjective questionnaires, BIARE and ENVIPE, which intersect in their fully representative state-wide applications in 2014, but also from traditional objective indicator sources such as the HDI and conventional surveys. This study uses two approaches, a descriptive analysis of a state-by-state landscape of indicators, both subjective and objective, in an initial search for stand-out well-being patterns, and an econometric study of a large selection of mainly subjective indicators inspired by theory and the findings of previous Mexican research. Descriptive analysis confirms that subjective well-being correlates strongly with and complements objective data, providing interesting directions for analysis. The econometrics literature indicates that happiness increases with income and satisfying of material needs as theory suggests, but also that Mexicans are relatively happy considering their mediocre incomes and high levels of insecurity, the last of which, by categorizing according to satisfaction with life, can be shown to impact poorer people disproportionately. The article suggests that well-being is a complex, multidimensional construct which can be revealed by using exploratory multi-regression and partial correlations models which juxtapose subjective and objective indicators.
\end{abstract}

\section{Introduction}\label{sec:intro}
Happiness may be the main objective of human existence. According to Greek philosopher Aristotle, all human behavior aims to achieve happiness. (\cite{Machado2015}). Research by \cite{Diener2013}  show that the vast majority of college students around the world consider happiness and life satisfaction to be extremely important and more important than money. This and other studies with similar conclusions justify researching and improving our measurement of well-being and quality of life (QoL) associated with the elusive concept of happiness. We know that well-being and QoL are complex subjects. ``[As] Seneca pointed out…reaching happiness is a challenging task...'' (\cite{Machado2015}, pp 100). A word of caution is perhaps in order, as the meanings of Subjective Well-Being (SWB), QoL and happiness certainly overlap but they are not perfect synonyms (\cite{Medvedev2018}). According to \cite{Steptoe2015}, SWB can mean three things--  evaluative well-being (or life satisfaction), hedonic well-being (feelings of happiness, sadness, anger, stress, and pain), and eudemonic well-being (sense of purpose and meaning in life), and that is the ample interpretation of SWB used here.\\

In attempts to improve understanding of the human condition, Objective Well-Being (OWB) expanded from its origins in income and expenditure to include approaches such as Basic Needs, Capabilities and Freedoms, while SWB measures satisfaction, the presence of positive feelings and the absence of negative ones. Using SWB is a direct way to complement and extend what is measured more indirectly with conventional OWB indicators like income, health and education. \cite{Diener2013} opine that researchers now know a lot more about what correlates with different aspects of well-being, both objective and subjective, so that it is more important  to understand the processes that underlie happiness, in terms of people's higher order goals, coping efforts, and dispositions.  \cite{Kubiszewski2019}  suggest that if we could know how to attain happiness, an important policy goal should be the creation of the conditions that allow people to maximize it. According to \cite{Angner2011} , pp24-25 ``The hope is that national well-being accounts will help policymakers understand and promote that which really matters...''. \\

\subsection{Objectives and Limitations}\label{subs:objectives}
The article examines the history of well-being and quality of life indicators, both objective and subjective, and the relationship among them. It uses available Mexican data to test how SWB can corroborate or complement the objective indicator story of well-being and, at the same time, validate recent findings concerning well-being in Mexico and Latin American. The most complete SWB data set for Mexico, BIARE abbreviated from \textit{Bienestar Auto Reportado} or Self-Reported Well-being, is fully representative at state level in the extended format only for 2014, while for 2013 there is a pilot survey and from 2015 to 2018, the basic survey application is representative solely at national level. A more narrowly focused SWB data set ENVIPE, abbreviated from \textit{Encuesta Nacional de Victimizacion y percepcion sobre Seguridad Publica} or National Survey of Victimization and Perception of Public Security, is representative at state level and available from 2011/12 through to 2018, but only cover the perception of security and related issues. Comprehensive OWB data from various censuses are available for periodic applications and fortunately most report for 2014, namely: 1) The \textit{Proyecto de Desarrollo de las Naciones Unidas} (PDNU) or United Nations Development Project (UNDP) website 2) The INEGI webpage Por \textit{Entidad Federativa} (i.e. by state) which combines state level indicators from different sources; 3) The \textit{Encuesta Nacional sobre Disponibilidad y Uso de Tecnologias de la Informacion en los Hogares} (ENDUTIH), 2015, which deals with the availability and use of household technology; and 4) The \textit{Consejo Nacional de la Evaluacion de la Politica Social} (CONEVAL), web site for social policy and program data. Consequently, 2014 is the reference year for the research. Most of the SWB indicators are presented as means, but also in population categories which report the intensity of replies, which is however not the case with OWB indicators, which are presented only as simple means with the information loss which that entails. \\

After this introduction of definitions, objectives and limitations, Section \ref{sec:history} analyzes the historical context of well-being indicators, both objective and subjective, underlining the importance of the Fitoussi Report. Section \ref{sec:recentfindings} introduces recent research from the well-being literature on issues including income, health, education, etc., with emphasis on the Mexican and Latin American experience. Section \ref{sec:methodologies} looks at the data sets and analytical options, in order to probe descriptively and quantitatively Mexican survey data to test postulations proposed in recent regional research before introducing a novel approach using partial correlation. Section \ref{sec:conclusion} concludes.

%--------------------------------------------------------
\section{Brief History of OWB and SWB}\label{sec:history}
%--------------------------------------------------------
Economics inspired OWB, the limitations of which are one reason for the emergence of the broader social sciences SWB, initially provoking a divergence between objectively and subjectively inspired well-being research, but their re-convergence in reports such as Fitoussi and subsequent research, is a pragmatic response to the need to better understand well-being, the conditions of individuals and their communities, and consequently improve public policy in areas such as welfare and urban planning. 
%--------------------------------------------------------
\subsection{OWB}\label{subs:owb}
%--------------------------------------------------------
Economic analysis has traditionally used OWB indicators for formal analysis. Conventionally, economic science tells (or should tell) us what ``is'', while normative and ethical inquires tell us what ``ought to be'' (\cite{Hands2012}, \cite{Robbins1952}) The distinction was made explicit by John Neville Keynes (father of John Maynard Keynes) who wrote: ``...whether political economy is to be regarded as a positive science, or as a normative science, or as an art, or as some combination of these, is...important...'' (\cite{Keynes2017}, pp. 22). Economists such as \cite{Samuelson2009} and  \cite{Friedman1953} later affirmed that ``positive'' economics was essentially a science which requires objectivity. According to \cite{Hands2012}, pp4, economic orthodoxy: required a ``...dichotomy -- the strict separation of the positive and the normative...[so]...According to logical positivism there were only two types of meaningful discourse -- empirical science (synthetic knowledge) and logic/mathematics (analytic knowledge) -- everything else was meaningless metaphysics.'' \\

In the 19th century, Economics and later sub-disciplines, such as Welfare Economics used ``utility'' to formalize the concept of well-being. A subsequent utilitarian argument – endorsed in the 20th century by Marshall, Pigou, and others – maintained that the diminishing marginal utility of money income provided Economics with scientific grounds for redistributing income from the richer to the poorer via taxation. Other economists such as Robbins considered interpersonal utility comparisons normative, preferring ``Pareto Efficiency'' as a policy maxim, which improves when at least one person benefits from an intervention without prejudicing anyone else in the community (\cite{Hands2012}). Later however, Welfare Economics pointed to incomplete information, imperfect markets and other distortions as justifications for improving societal welfare government via policies of redistribution (\cite{Greenwald1986}, \cite{Stiglitz1991}).\\

From the middle of the 20th Century Development Economists refuted the conventional wisdom which confines Economics to rigorous objectivity, for example, \cite{Myrdal1958}, who argued that  ``...normative values are inexorably intertwined with economic science...'' (\cite{Hands2012}, pp7). So, since the Second World War, objective indicators have been used to measure well-being, and up to the 1960s, Economists proxied OWB by measuring household income and expenditure. In the 1970s a rather materially orientated Basic Needs multidimensional approach combined measures of household inadequacies concerning absent amenities and services, overcrowding, inter-household economic dependence and non-attendance at school (\cite{ILPES2001}, \cite{Singh1979}, \cite{Streeten1981}). In the 1980s and 1990s, Sen's initially abstract Capabilities and Freedoms approach was operationalized by various authors, including philosopher Nussbaum, who mapped out 10 basic human capacities (\cite{Nussbaum2000}, \cite{Sen1981} \cite{Sen1985}, \cite{Sen1999}). Independently but at the same time, \cite{Doyal1991}, also philosophers, produced something similar in their Theory of Human Need. Sen, working with a group of economists at the United Nations Development Project (UNDP), published a first version of the Human Development Index (HDI) in the  Human Development Report (HDR), 1990, as a simple average of three indicators, namely life expectancy, years of schooling and income per capita, and as such is a reductionist vision of the Capacities approach. Since then, the HDR has experimented with large numbers of well-being indicators, although the HDI still combines three indicators for health, education and income albeit in a reformed formulation since 2010.\\
 
%--------------------------------------------------------
\subsection{SWB}\label{subs:swb}
%--------------------------------------------------------
SWB has been studied using well-being indicators through most of the 20th century in the fields of Psychology, Sociology and Education, rather than Economics (\cite{Angner2011}). \cite{Porter1995}  mentions that subjective measures were used in marital success studies, Educational Psychology, Personality Psychology, Epidemiology of mental health and Gerontology from the 1920s and 1930s onwards. In the 1960s and 1970s the ``Social Indicator Movement'' spread through the disciplines of Sociology and Education (\cite{Bauer1966}), leading to the launch of the Social Indicators Research Journal in 1974. Initially, social indicators were overwhelmingly objective, and according to \cite{Porter1995}  the quantification of well-being ``made up'' for a lack of trust in ``soft'' disciplines like psychology. However, from the 1970's onward, subjective indicators were increasingly incorporated into well-being studies. \cite{Solomon1980}  reserved the term ``social indicators'' for what we now call ``objective indicators'' and ``quality-of-life indicators'' for what we now call ``subjective indicators'', however these have not become standard definitions (\cite{Land2018}). The ``Social Indicator Movement'' was motivated by a desire to develop accurate measures of the quality or goodness of life (\cite{Andrews1989}, \cite{Campbell1976}, believing that widely used economic measures were imperfect proxies for well-being, and proposed that ``...[the USA] ...change...its fixation on goals which are basically economic to goals which are essentially psychological..'' (\cite{Angner2011}, pp24). A measure of progress since the beginning of the ``Social Indicator Movement'' is the existence of a number of journals which publish research on SWB, such as: Social Indicators Research, Quality of Life Research, Journal of Happiness Studies, Applied Research in Quality of Life, and Child Indicators Research. 
%--------------------------------------------------------
\subsection{The Stiglitz, Sen and Fitoussi Report}\label{subs:SSFreport}
%--------------------------------------------------------
The report by the Commission on the Measurement of Economic Performance and Social Progress by Stiglitz, Sen and Fitoussi  \cite{Stiglitz2009}, popularly called the Fitoussi Report, lent credence to operationalization of SWB. The commission was created at the beginning of 2008 on the initiative of the French government with the objective of identifying the limits of Gross Domestic Product (GDP) as an indicator of economic performance and social progress, signaling a shift of emphasis from measuring economic production to measuring people's well-being and also in the context of environmental sustainability (\cite{Stiglitz2009}). Universally, standardized GDP is still the most widely used objective measure of economic activity, but it measures mainly market production rather than well-being. Conflating the two mixes up ``well off'' with ``well-being'' and can encourage the implementation of erroneous polices, for example orientated towards economic growth. According to this argument, complementary measures of both OWB and SWB provide key information about people's QoL, and government statistical offices need to incorporate surveys and questions to capture people's  perceptions  and  priorities (\cite{Stiglitz2009}).\\

Three approaches to well-being are presented in the Fitoussi report, two in the OWB tradition, namely: 1) The Fair Allocation approach from Welfare Economics (which is where Sen started out as an economist); and 2) The Capabilities approach (later developed by Sen). The third is qualitatively distinct: 3) SWB approach. According to \cite{Stiglitz2009}, the three approaches have obvious differences and certain similarities and the choice between these approaches is ultimately a normative decision. Cultural heterogeneity problems outstanding, the three approaches mesh into a multidimensional definition of well-being which can be constructed round the following eight dimensions: 1) Material living standards 2) Health 3) Education 4) Personal activities (including work and leisure) 5) Political voice and governance 6) Social connections and relationships 7) Environment and 8) Insecurity, of an economic as well as a physical nature. Finally, rather than a dimension, inequality is considered a cross-cutting issue of importance. \\

The Fitoussi report makes a series of recommendations for future well-being studies which have influenced subsequent studies in many countries, including INEGI's BIARE survey. A methodology document published by INEGI explains ``The conceptual reference and starting point of these ideas were established in the Stiglitz-Sen-Fitoussi commission by the French government in…2009...'' (\cite{INEGI2015}, pp1). In other words, INEGI expressly used the Stiglitz report to justify the BIARE approach to measuring well-being in Mexico. 

%-----------------------------------------------------
\section{Recent Findings in the Well-being Literature}\label{sec:recentfindings}
%--------------------------------------------------------
Recent international empirical studies of well-being in specialized journals have thrown up some interesting and frequently contradictory findings on a number of issues, including material living standards and inequality, health and education, individualism and community, the multidimensionality of well-being and the paradox of Latin American happiness in the face of adversity.
%--------------------------------------------------------
\subsection{Material living standards and inequality}\label{subs:materialliving}
%--------------------------------------------------------
According to \cite{Keny2005}, over a wide sample of countries, both developing and rich, objective QoL indicators are positively related to income, so that people in poor countries are unhappier than people in rich countries. However, this positive connection is stronger in poorer developing countries. In Australian data a law of diminishing returns appears to operate for SWB and for income above an empirically defined threshold (\cite{Cummins2018}), which is corroborated by a study from the USA, in which rising incomes yield diminishing returns to happiness (\cite{Eckersley2009}). Another study reports a positive association between needs satisfaction and SWB in relatively poor Bangladesh, but the same does not occur in Malaysia, where superior incomes apparently encourage Malaysians to compare themselves with wealthier neighbors, undermining their perception of self-improvement (\cite{Camfield2010}). Research in Peru finds an apparent mismatch between perceptions of material evidence of well-being, where rural inhabitants are poor but happy, while those that move to urban areas motivated by self-improvement find it difficult to satisfy heightened aspirations (\cite{Copestake2007}). Likewise, in Brazil, perceived income sufficiency depends positively on absolute family income and negatively on relative neighborly income (\cite{GoriMaia2013}). A comparable conclusion is reached by a South Africa Cape Area Panel Study on the SWB of young adults, which appears to depend not only on the comparisons they make with those around them, but also with themselves across time (\cite{Tibesigwa2016}). So, it appears that income is important to well-being, but that does not make it a proxy of it, because other life domains are important too ( \cite{Rojas2004}). In the words of \cite{Rojas2018}, reviewing regional data, ``...the Latin American case does not ignore the importance of income, but it clearly shows that there is more to life than income''. \\

A cross-cutting issue is inequality and it is interesting that a cross-country review by  \cite{Evans2019} finds that income improvements increase SWB whereas inequality appears to be irrelevant, so that a decrease in the Gini coefficient does not correlate with an increase in SWB at the individual level. On the other hand, in Brazil, according to \cite{GoriMaia2013}, improving personal income and education, as well as reducing inequality, is the most effective way to improve QoL and perceived well-being in society, however there are obvious methodological issues at stake, like separating and discerning the effects of different indicators, and also whether the unit of study is the individual or the community, and how one affects the other. The inconclusiveness of inequality research is flagged by   \cite{Schneider2016}, who reports that it remains unclear whether people living in areas of high income-disparity feel better off or less well off than people living in environments where everyone is more equal, because different researchers in different countries report positive, negative or no apparent effects of inequality on well-being.
%--------------------------------------------------------
\subsection{The Basic Development Building Blocks of Health and Education}
\label{subs:basicdevelop}
%--------------------------------------------------------
The relationship between health status and subjective well-being has been extensively studied given health's pivotal position in life (\cite{Maccagnan2017}). Studies consistently reveal a strong relationship between health and happiness, stronger in fact than that between happiness and income. Health shocks -such as serious diseases or permanent disabilities- have negative and often lasting effects on happiness. At the same time, happier people are healthier, so causality runs in both directions (\cite{Graham2008}). However, the relation between health and happiness is not always close. Healthy people are not uniformly happy and people with severe health limitations or who suffer accidents can adjust to their circumstances and report or recover normal levels of happiness (\cite{Subramanian2005}). There are several pathways through which subjective well-being can affect health, the first being that happier individuals have a healthier lifestyle, with respect to smoking, exercise, healthy eating, sleep habits and adherence to treatment regimes. Well-being also appears to affect health physiologically, provoking better neuroendocrine, inflammatory, and cardiovascular responses. Third, positive emotions can offset negative ones (\cite{Maccagnan2017}). Poverty in low- and middle-income countries has adverse effects on health, with low educational levels and low social status associated with higher prevalence of mental illness, including depression and anxiety (\cite{Cazzuffi2018}). Health also affects social life. Significant differences were found between two groups of Algerians who self-categorized themselves as healthy and unhealthy, with respect to pain, anxiety and sleep habits. The healthier group showed significantly higher satisfaction with marriage, friendship and family relationships, suggesting a positive relation between objectively defined good health and social relationships (\cite{Tiliouine2009}). Extending parameters, a Mexican study finds a significant relation between lowly individual psychosocial well-being (i.e. self-reported depression symptoms, feelings of sadness and experience of stress) and poor scores for objective house-hold indicators (income, wealth, number of residents, employment status and dependency), deprivation at municipal level (head count poverty, violence (specifically murders) and income inequality (measured by the Gini coefficient) (\cite{Cazzuffi2018}). \\

Many studies and in different countries find that education and SWB are positively related, at least in descriptive analysis (\cite{Maccagnan2017}), however where econometric studies do not control for income and wealth, their impact may be transmitted within the education variable, risking the attribution of spurious relations. It is also amply demonstrated that income returns to education are universally high, often accompanied with improvements in self-esteem, success in marrying, but also a loss of leisure time and an increase in stress (\cite{McMahon2009}). Where income and wealth variables are included in econometric studies, the relationship between education and SWB can even turn negative, which has been explained by rising but unfulfilled expectations created by a better education; not so dissimilar to the relative income hypothesis (refer to Section \ref{subs:materialliving}). Using Japanese survey data, \cite{Clark2015} show that much of the effect of education on SWB is cancelled out by increases in aspirations. In an Australian study, the effect of increased education on SWB turns out to be neutral, or more precisely it is postulated that people with higher levels of education do tend to have higher expectations of life, but do not systematically differ from the rest of the population in their ability to meet those expectations (\cite{Kristoffersen2018},  \cite{Clark2015}).
%--------------------------------------------------------
\subsection{Individualism, Community, and Westernization}\label{subs:individualism}
%--------------------------------------------------------
Echoing findings on inequality (refer to Section \ref{subs:materialliving}), it is assumed by many researchers that SWB is determined mainly by individual characteristics. The expectation is that specific features of communities or neighborhoods do have an impact on the QoL of citizens, but little is known about exactly which ones (\cite{Hooghe2011}). According to \cite{Helliwell2004}  ``Social capital is strongly linked to SWB through many independent channels and in several different forms. Marriage and family, ties to friends and neighbors, workplace ties, civic engagement (both individually and collectively), trustworthiness and trust: all appear independently and robustly related to happiness and life satisfaction''.  South Korean research corroborates and extends the connection, postulating that the natural environment, local administration, and social capital are the most important factors for overall perceived QoL in Korean cities. Accordingly, resident satisfaction with collective goods, rather than private goods, shape overall assessments of well-being in the community. That being the case, scholars and policymakers should pay more attention to collective goods that enhance community well-being (\cite{Lee2018}). \\

An interesting reflection is whether well-being indicators confuse QoL with individualism and westernization?  \cite{Diener2013} present a strong argument in favor of using SWB indicators, however they typically measure individual well-being, rather than the societal well-being. Asking people about life in general or the lives of others gives a very different result from asking them about their own lives. Interviewing picks this distinction up, because interviewees downgrade their individual well-being if they have previously answered questions on societal or political issues. Researchers tend to buffer this ``contamination'' by using interview devices to “change the subject”, as if only individual well-being matters (\cite{Eckersley2009}, \cite{Eckersley2013}). Studies may have purposefully or inadvertently left out societal issues which link back into individual well-being in ways which have not been detected or understood. So, for example, in a study of Finnish youth, mental issues such as fears of loneliness or suicide have been increasing in recent decades, but alongside measured improvements in well-being, which means that something is not being correctly identified and very likely we do not understand the pathways linking individual and societal well-being. (\cite{Lindfors2012}). Thus, \cite{Eckersley2013} in a cross-country review, suggests that SWB as it has been defined by scholars may measure individualism, modernization and westernization rather than improved QoL or human progress. In more general terms, this argument is important, because it points out that OWB indicators can be biased both in their initial definition (because we impute their importance as proxies of well-being) and in their application (i.e. how we survey), a point made by \cite{Rojas2011} in defense of the SWB approach (refer to Section \ref{subs:multidim} below).
%--------------------------------------------------------
\subsection{Multidimensionality of Well-being}\label{subs:multidim}
%--------------------------------------------------------
\cite{Fernandez2019}  discuss the multidimensional nature of well-being using PCA to categorize indicators in dimensions, such as material well-being, health, education, etc. and derive weights from PCA output  which indicate their relative importance in explaining variation. Multidimensional life satisfaction studies by Rojas using Mexican data similarly identify a number of domains using PCA, including: health, economic, job, family, friendship, personal and community, in order to create composite indexes, and use econometric modelling to determine the relative importance of those domains (\cite{Rojas2006}). According to \cite{Fernandez2019}, OWB indicators describe most of the variation in Mexican well-being, which is strongly heterogeneous geographically across the 32 States. Interestingly, existing indexes, both objective and subjective, rank the States in similar orders according to relative well-being, which means that the subjective corroborates the objective, or visa-versa. The principal sources of indicators considered include INEGI objective indicator surveys and the BIARE application for subjective indicators. According to  \cite{Rojas2004}, \cite{Rojas2011}, SWB is invaluable because it measures life satisfaction as affirmed by the people themselves, dealing with human beings as ``flesh and blood'' (\cite{Rojas2004}) rather than what is presumed or imputed to be important by researchers using objective indicators like income, levels of literacy or access to social amenities, which can suffer from errors of misspecification (\cite{Rojas2011} -- refer to Section \ref{subs:individualism}). Internationally, studies experimenting with SWB indicators since the 1970's believe they not only complement objective indicators but open up new avenues for understanding well-being and QoL, achieving true multidimensionally in the process (\cite{Hallerod2013}, \cite{Kubiszewski2019}, \cite{Oswald2010}). \cite{Fernandez2019} recommend that the multidimensional nature of well-being be incorporated into public policy to avoid the dangers of misinterpretation due to partial and biased measures of the phenomenon. 
%--------------------------------------------------------
\subsection{The Paradox of Latin American Happiness in the Face of Adversity}\label{subs:paradox}
%--------------------------------------------------------
An interesting finding is the detection of relatively high levels of well-being created by close extended family networks which are an essential part of a Latin American lifestyle and appear to more than compensate for relatively low incomes, pervasive poverty, inequality  crime and violence. In other words, income is significantly relevant to well-being as are many factors, but a closely-knit, active, family life is probably more important and in Mexico and Costa Rica, warm, and person-based family relations substantially contribute to happiness (\cite{Rojas2012}, \cite{Rojas2018}). However, Latin Americans are not immune to their many social and economic problems. For example, according to an econometric analysis by \cite{Rojas2018}, satisfaction with life declines in the presence of perceptions of corruption, exposure to crime and economic difficulties.\\

According to \cite{Martinez2018}, ``...there is a dearth in evidence at the state level on how public-insecurity indicators interact or are included in a multidimensional measurement of well-being'' (pp.454). They build multidimensional indexes for material welfare, economic well-being, SWB, social capital and public insecurity, using INEGI sources of objective indicators, but also, BIARE and ENVIPE sources. They use a regression method to create synthetic indices to measure change patterns of violence in Mexico over time (although information sources are cross-sectional). According to them, violence has evolved from a largely rural southern phenomenon towards an urban reality linked to organized crime throughout the country. The authors maintain that strong social ties and networks help foster a sense of security in the face of rampant crime, and that perceptions of public insecurity can also be interpreted as a breakdown of social networks. They report that multidimensional indexes can readily use available INEGI surveys, although the design of a new, single but wide ranging and holistic survey could better identify the importance of crime, violence and public security in the well-being and QoL equation.  \cite{Charles2018} using the BIARE INEGI data with linear regression analysis, show that violence negatively influences SWB, and the effect is greater for woman. On the other hand, social and family networks reduce the likelihood of physical assaults, threats and violence, although speaking an indigenous language increases it. Interestingly, women who participate in the labor force are not less likely to suffer domestic violence than those who do not. Education is important, as the more education a woman has, the less violence inflicted on her, so in terms of public policy, social programs could work to improve gender equity through social networks and education (\cite{Charles2018}). Not everything is a problem in Latin America, and public policy needs to recognize the family and community as a source of strength and inspiration (\cite{Rojas2012}, \cite{Rojas2018}).

%--------------------------------------------------------
\section{Methodologies and Results}\label{sec:methodologies}
%--------------------------------------------------------
The descriptive analysis and econometric models use SWB indicators from two sources, BIARE and ENVIPE, and OWB indicators from the UNDP and various INEGI surveys (refer to Section \ref{subs:objectives}). The data is freely available on the internet.

%--------------------------------------------------------
\subsection{Descriptive Analysis}\label{subs:descripAnl}
%--------------------------------------------------------
For descriptive analysis, the BIARE and ENVIPE and various objective indicator census sources (refer to Objectives and Limitations) are uploaded onto a symmetrical spreadsheet format, state by state, with 39 SWB BIARE indicators and 29 SWB ENVIPE  indicators, making a grand total of 68 SWB indicators, arranged as per their original survey domains with 88 OWB indicators from various census sources grouped into 13 dimension or domains, namely: population, multidimensional well-being; income, inequality, education, health, social security, gender issues, work, household amenities and services, crime and security, rurality and agriculture, and availability and use of household technology. Some of the indicators overlap, for example multidimensional poverty indicators, by definition, combine individual indicators, while indicators from different surveys can be almost identical. The only requirement here is that they are reputable and representative at federal entity (State level); and in fact, all of them are sourced from INEGI. The arrangement enables the application of Pearson correlations through the data set of over 150 indicators, a panoramic view Mexican well-being. \\

\begin{figure}[H]
	\begin{center}
		\includegraphics[width=5cm]{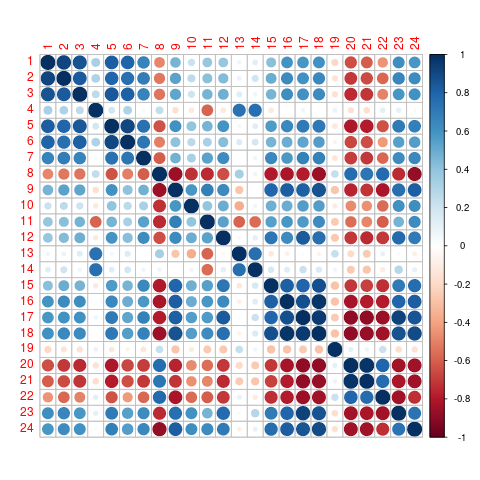}
	\end{center}
	\caption{Correlogram of Selected Subjective and Objective Indicators for Mexico, 2014. Keys: 1. Life in general
		2. Health
		3. Achievements
		4. Satisfaction with personal security
		5. Material needs satisfied
		6. Excellent living conditions
		7. Wouldn't change anything if born again
		8. Perception of poverty
		9. Perception of corruption
		10. Criminals go unpunished
		11. Insecurity
		12. Narcotraffick
		13. Perception of local security 
		14. Perception of municipal security
		15. Health Indicator IDH
		16. Education Indicator IDH
		17. Income Indicator IDH
		18. IDH
		19. GINI 
		20. \% pop. with at least one social deprivation
		21. \% pop. without access to social security
		22. \% pop.in rural localities
		23. Households with access to internet
		24. Salaried employment.}
	\label{FigCorrelogram}
\end{figure}

A selection of indicators from the spreadsheet is presented in Figure \ref{FigCorrelogram} and captures the key features of the data correlations. Note that the correlations can be strongly positive or negative, depending on the original specification of the indicators, for example, perceptions of both security and insecurity crop up in different sections of the ENVIPE survey. That perhaps is an extreme case, however, the phenomenon is ubiquitous, and another example is indicators for both informal and salaried employment. In Figure \ref{FigCorrelogram} we choose to present salaried employment, in an attempt overall to represent both development and underdevelopment issues, and we obtain both positive and negative correlations, depending on the framing of the other indicators. If we had included the indicator informal employment, which is in our database, the correlations are equally strong, but have the opposite signs. The selected indicators come from the BIARE survey (indicators 1 through to 7), the ENVIPE survey (indicators 8 through to 14), UNDP data (indicators 15 through 18) and objective indicators from conventional surveys (refer to Objectives and Limitations Subsection \ref{subs:objectives}).\\
Comparing OWB with OWB indicators for the 32 Mexican states is not our research objective but it provides background and tells a similar story to previous studies, which is that OWB indicators strongly correlate with each other in geographical analysis, due to decades long processes of unequal development with cumulative causation (Lewis, Myrdal, Myint, Krugman, etc.). (\cite{Heald2018}). There are strong correlations across indicators for multidimensional well-being, incomes, education, household quality and amenities, household access to internet and ownership of computers, rurality (negatively) and salaried work. We can see this in Figure \ref{FigCorrelogram} by observing the bottom right quadrant. However, the Gini coefficient which measures inequality correlates poorly with the rest of the indicator landscape, as it does in other studies (\cite{Heald2018}), underlining its ambiguous impact on well-being (refer to Section \ref{subs:materialliving}). Interestingly, when we compare SWB with OWB indicators (refer to the top-right quadrant in Figure \ref{FigCorrelogram}), and starting with the BIARE survey summary indicator ``Satisfaction with life in general'', we again obtain strong correlations with the same OWB indicator domains, be the multidimensional well-being, income, etc., and again a poor correlation with the Gini coefficient. This suggests a deep relationship between subjective and objective findings, validating the Fitoussi recommendations for Mexico at the level of States on including SWB in QoL studies (refer to Section \ref{subs:SSFreport}). In general, over half of the BIARE subjective indicators share similar correlation patterns to the summary indicator i.e. correlate widely across the indicator landscape, although the rest do not (and for which there is no simple explanation). Concerning the ENVIPE survey, which specializes in crime and security perceptions, just five of the indicators correlate strongly with our universe of objective indicators, and the rest do not, which is perhaps not surprising considering the specialist nature of the survey which intersects less generally across indicator domains. In Figure \ref{FigCorrelogram}, we show various ENVIPE indicators (8 to 12) which do correlate strongly.\\

Comparing SWB with SWB indicators in the top, left quadrant, strong correlations exist within surveys, for example BIARE with BIARE indicators (1 to 7 in Figure \ref{FigCorrelogram}), or ENVIPE with ENVIPE indicators (9 to 14), which is expected considering that the questions are based on perceptions which can overlap. More intriguing are correlations between BIARE and ENVIPE indicators. Interestingly, the BIARE perception of personal security correlates with the ENVIPE indicators for perceived insecurity, and perception of security at local and municipal level (refer to Figure \ref{FigCorrelogram} – indicators 4, 11, 13 and 14, respectively). We put that into perspective by noting that indicators 4, 13 and 14 all correlate very poorly with everything else. That two SWB surveys applied independently corroborate perceptions of crime, violence and insecurity is revealing, strengthening the conviction that well-constructed SWB surveys are reliable and useful. \\

The correlation analysis broadly confirms recent literature findings on well-being, namely that it is multidimensional and complex and that SWB indicators corroborate and complement both themselves and OWB indicators, and visa-versa (refer to Section  \ref{sec:recentfindings}). It also sets the stage for some more in-depth econometric analysis (refer to Sections \ref{subs:quantitativeAnalysis}- \ref{subsec:ggmmicrodata})
%--------------------------------------------------------
\subsection{Quantitative Analysis}\label{subs:quantitativeAnalysis}
%--------------------------------------------------------
For statistical analysis, instead of analyzing over 150 variables simultaneously as in Section \ref{subs:descripAnl}, here we follow a deductive procedure  based on  theory-based models, of a small number of indicators which best explain the variation in the data landscape. This pathway moves from a literature survey of theory and previous findings, through observation, validation or the emergence of new hypotheses, which is the approach used here. Linear regression and logistic models using odds ratios can use means or population data grouped into categories, for dependent or independent variables. Categorized population data can analyze relationships at opposite ends of indicator ranges which enables the testing of more sophisticated hypotheses (refer to Objectives and Limitations Subsection \ref{subs:objectives}).\\

For the econometric models in Section \ref{subsec:aggegateddata} we use aggregated data, or more especifically, weighted averages for each of the the 32 States, which allows to combine information from different sources, such as BIARE, ENVIPE and PNUD.   Only one question is selected from the ENVIPE survey, namely $AP4-3-3$:  ``In terms of criminality, do you consider the State you live in to be secure / insecure?''. Concerning OWB indicators only the income component of the HDI is used, as reported by UNDP Mexico.\\

In Subsections \ref{subsec:logisticmicrodata} and \ref{subsec:ggmmicrodata} we use a micro-data approach by selecting information exclusively sourced from BIARE.  We take two blocks of BIARE questions. Table  \ref{table:biare1} in Appendix \ref{appendix:blocks} lists 16 questions, each graded from 0 to 10, with 10 denoting maximum satisfaction. In Table \ref{table:biare2} a second block is listed, each question graded from 1 to 7, also from BIARE.

%-------------------------------------------------------
\subsection{Aggregated data}\label{subsec:aggegateddata}
\begin{figure}[H]
\begin{center}
\includegraphics[width=10cm,height=5cm]{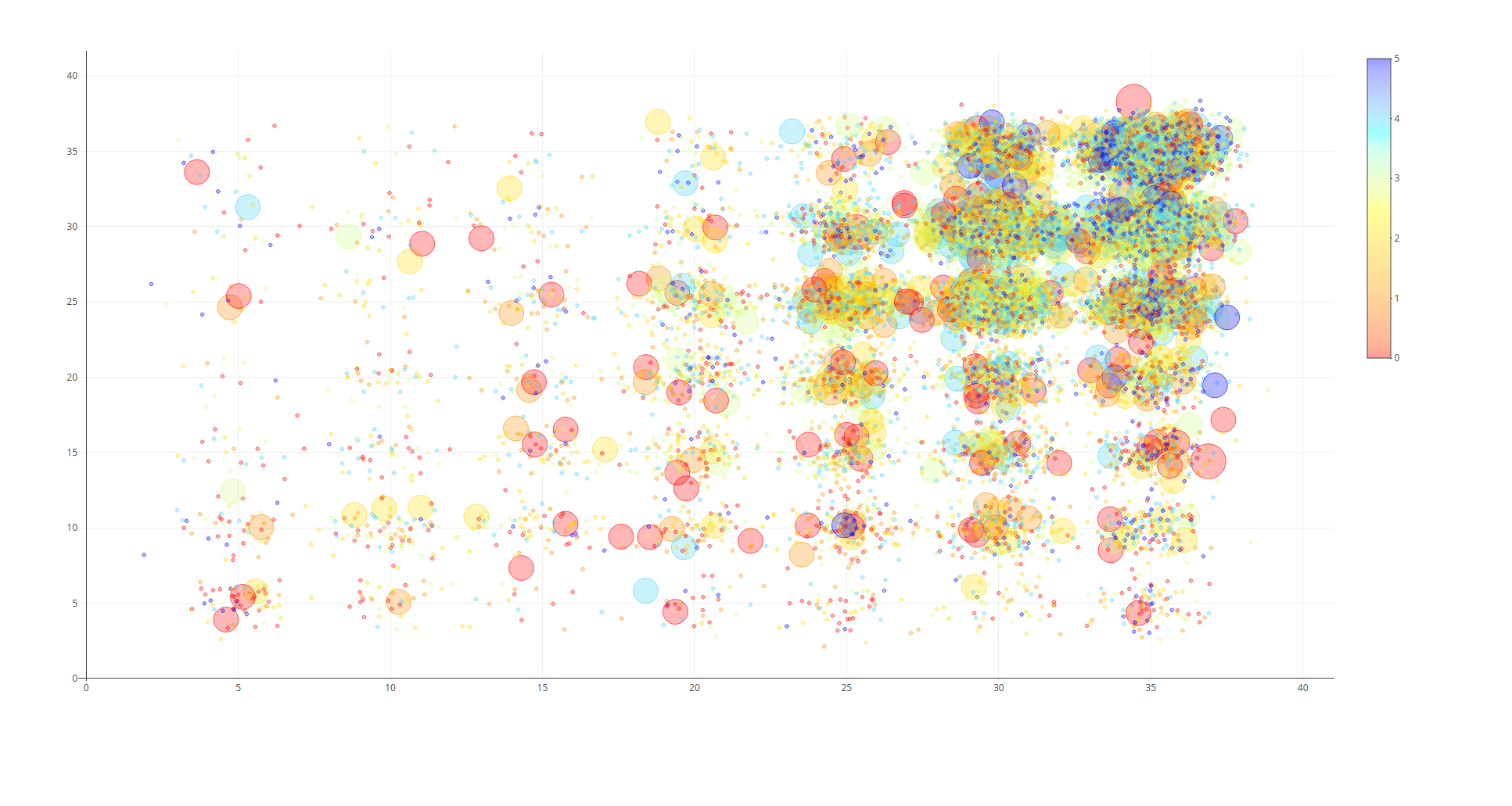}
\end{center}
	\caption{Bubble Plot of the Relation between Happiness and the Satisfaction of Material Needs, with Personal Security Superimposed.}
	\label{FigBubbleplot}
\end{figure}

%--------------------------------------------------------
Section   \ref{subs:materialliving} discussed the importance of income, Section  \ref{subs:multidim} the multidimensional nature of well-being, and Section \ref{subs:paradox} the paradox of happiness in Mexico and elsewhere in the region but accompanied with high levels of poverty and violence. At the national level, the BIARE data set has 39,274 observations. As an initial inspection of our data, we select three variables to analyze the interplay between SWB, in this case hedonic  happiness, and OWB, represented here by an evaluative SWB indicator of material needs satisfaction. Mexico, like other parts of Latin America has suffered years of violence, in Mexico's case due to drug trafficking, so we are also interested in discovering how the perception of personal security links into well-being.  In Figure \ref{FigBubbleplot}, 50\% of these observations, selected at random, are displayed with the following interpretation: 
\begin{enumerate}
\item  The color of the circle is an indicator of citizen perception, with blue representing a positive perception of personal security and red the opposite (Question $satis_9$).
\item The size of the circle represents population size according to the survey design. 
\item The $y$ axis represents satisfaction concerning covering material needs of household members (Question $afirma_2$)
\item The $x$ axis represents satisfaction concerning personal happiness (Question $afirma_1$).
\end{enumerate}
Figure \ref{FigBubbleplot} illustrates a positive relation between the hedonic SWB question of  happiness perception ($afirma_1$: ``The household member is a happy person'') and in satisfying material needs of household members ($afirma_2$: ``The key household material needs of household members are satisfied''). This means that from an individual perspective there is a relationship between evaluative SWB (measured as an individual´s success in satisfying material needs) and hedonic SWB (measured as an individual´s expressed happiness with life). On the other hand, the color of the circles indicates how the individual scores her or his personal security (Question $satis_9$: ``How satisfied are you concerning your personal security?''). It is evident that the red and yellow colors predominate indicating a generalized perception of poor personal security. However, it is notable that despite a negative generalized perception of insecurity, many individuals reveal high levels of expressed happiness, which concurs with the Latin American well-being literature (refer to Section \ref{subs:paradox}). On the other hand, the graph also shows individuals with a positive perception of security occupying the top, right hand side of the graph, where higher levels of happiness, material needs, and personal security are satisfied and suggests that insecurity disproportionally effects poorer families. This suggests that well-being is multidimensional (refer to Section \ref{subs:multidim}) as evidenced through mutually reinforcing indicators. This visual evidence will be discussed and confirmed below via the adjustment of a set of statistical models.\\

To give a more quantitative description of Figure \ref{FigBubbleplot} beyond visual evidence, we start with some basic linear regressions which provide some interesting initial results concerning the multidimensional nature of well-being (refer to Section \ref{subs:multidim}) using the BIARE, ENVIPE and UNDP Mexico \cite{UNDP2016} data. We then develop non-linear models using logistic regression and a Gaussian graphical model in Subsection  \ref{subsec:ggmmicrodata} below.
\begin{figure}[H]
	\begin{minipage}{.5\textwidth}
		\centering
		\includegraphics[width=.8\textwidth]{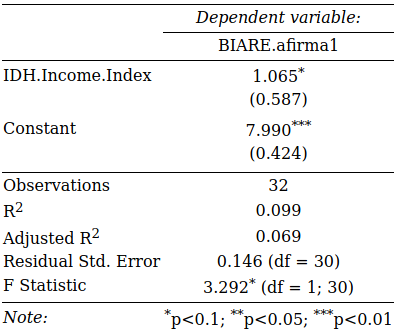}
		\subcaption[second caption.]{}\label{figre0}
	\end{minipage} 
	\begin{minipage}{.5\textwidth}
		\centering
		\includegraphics[width=.8\textwidth]{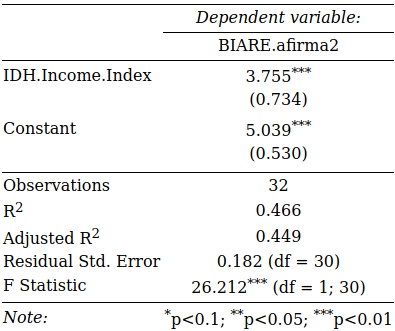}
		\subcaption[second caption.]{}\label{figre00}
	\end{minipage}
	\caption{Results of linear regressions.}\label{figre00Full}
\end{figure}
The results of a linear regression of aggregated values (weighted averages for each federal entity, i.e. state) are displayed in Figure  \ref{figre00Full}. In the left panel \ref{figre0},  the dependent SWB variable corresponds to $afirma_1$: ``The household member is a happy person'' and the explanatory variable is the OWB weighted average of the income component of the Human Development Index (UNDP, 2014). Although significant and with the expected sign indicating a positive relationship, the fit of the model is poor. This confirms the literature findings reported in Section  \ref{subs:materialliving}, that income is linked to well-being, but so are other factors. In the right-hand side panel \ref{figre00}, the same explanatory variable is used, however the dependent variable corresponds to $afirma_2$ concerning: ``satisfying material needs of household members''. The fit of the model is clearly superior to the left hand one, which is to be expected owing to the obvious relation between objective income and satisfying material needs.\\

\begin{figure}[H]
	\begin{minipage}{.5\textwidth}
		\centering
		\includegraphics[width=.8\textwidth]{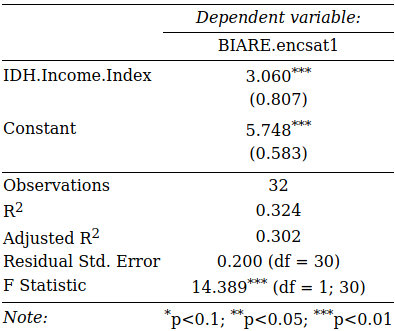}
		\subcaption[second caption.]{}\label{figre1}
	\end{minipage} 
	\begin{minipage}{.5\textwidth}
		\centering
		\includegraphics[width=.8\textwidth]{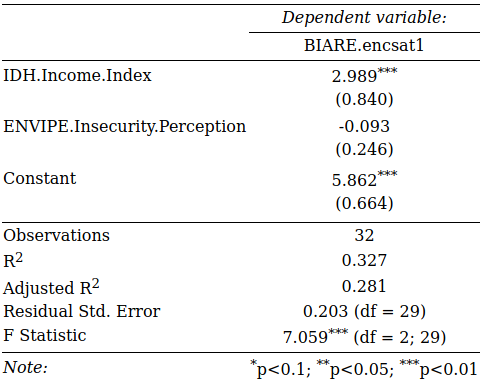}
		\subcaption[second caption.]{}\label{figre2}
	\end{minipage}
	\caption{Results of regressions.}\label{figre12Full}
\end{figure}

So far in the first regressions we have explored the relationship between a hedonic SWB indicator (happiness) with the income component of HDI, an OWB indicator, and in the second regression the relationship between an evaluative SWB indicator (success in satisfying material needs) also with the OWB income component of the HDI. In Figure \ref{figre12Full} we explore an alternative dependent variable.  In the left panel \ref{figre1}, is the result of a linear regression which presents the dependent variable as SWB indicator $encsat_1$: ``How satisfied are you with your current life?'' and the explanatory variable once more as the mean of the weighted average of income from the Human Development Index (UNDP  \cite{UNDP2016}). Again there is a significantly positive relationship between SWB and OWB indicators. The panel on the right \ref{figre2} illustrates an extended version of the model which also includes the weighted average perception of insecurity variable from the ENVIPE 2014 database  (question $AP4-3-3$ with text: ``¿In terms of criminality, do you consider the State you live in to be secure / insecure?''). It is notable that the coefficient of the insecurity variable has an expected negative sign, although it is not statistically significant. In 2014 the wealthier States like Mexico City, Nuevo Leon, and others in the north nearer the US frontier were involved in a violent drugs war. So, income impacts positively on happiness, but not violence and insecurity, which has the opposite effect. This finding is hardly surprising considering the evidence in Figure \ref{FigBubbleplot}, in which a  perception of insecurity predominates despite positive scores for both satisfaction of material needs and happiness. It suggests however that the perception of insecurity integrates in complex ways with other individual sentiments and perceptions, which requires a more sophisticated econometric model. This is tested below through  more sophisticated models as logistic regression and Gaussian graphical model revealing partial correlations (refer respectively to Sections \ref{subsec:logisticmicrodata} and  \ref{subsec:ggmmicrodata} below).\\
%--------------------------------------------------------
\subsection{Logistic regression with microdata}\label{subsec:logisticmicrodata}
\begin{figure}[H]
\begin{center}
\includegraphics[width=5cm]{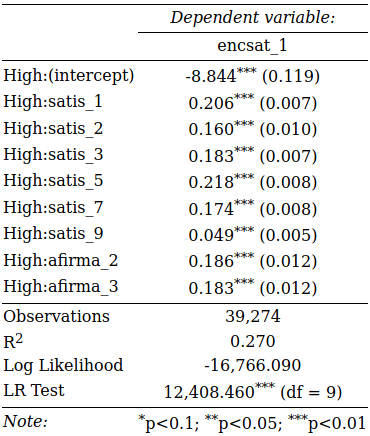}
\end{center}
\caption{The Relation of Satisfaction of Current Life with a Basket of Subjective Life Satisfaction Indicators. Logit model.}
\label{figmylogit}
\end{figure}

%--------------------------------------------------------
Figure \ref{figmylogit} shows the result of estimating a multinomial logistic regression; for a systematic presentation of this model see e.g., \cite{Hosmer2013}.  The dependent variable is BIARE $encsat_1$: ``How satisfied are you with your current life?'' and the following explanatory variables (refer to Table \ref{table:biare1}):  $satis_1$ (concerning Social life), $satis_2$ (Family life), $satis_3$ (Affective life), $satis_5$ (Health), $satis_7$ (Perspectives for the future), $satis_9$ (Perception of personal security). From Table \ref{table:biare2} we also include $afirma_2$ (Satisfaction of material necessities) and $afirma_3$ (life conditions of household member are excellent). There is an important difference with respect to the previous linear regressions in that here the estimation is carried out using data exclusively from the BIARE questionnaire and all the variables measure SWB. The variable from question $encsat_1$ ``How satisfied are you with your current life?'' is classified into two answer categories (which presented best fit and in which cases are in equilibrium), ``High'' (with an 8 to 10 score), and ``Low'' (0-7). If we use $Y$ to denote the variable which has a zero value except when Question $encsat_1$ is ``High'', and $X$ as a vector of explanatory variables, the proposed model takes the form:
\begin{multline}
logit \mathbb{P}[Y=1 \mid X]=a_1 + a_2 satis_1+a_3 satis_2 + a_4 satis_3 + a_5 satis_5\\
 + a_6 satis_7+ a_7 satis_9 + a_8 afirma_2 + a_9 afirma_3.
\end{multline}
The category ``Low'' is established as the reference. All the coefficients have the expected signs, and all are significant. For example, the coefficient of “Satisfaction with family life” is equal to 0.160, indicating that for a score of $X$ points for this concept, the probability of obtaining the response ``High'' to the question $encsat_1$, is proportional to: $e^{0.160 x}$, in which case the higher the score for the question concerning ``Family life'', the higher the probability of obtaining a high score for ``Satisfaction with current life''. The rest of the coefficients have similar interpretations. Note that the perception of security is now positively significant too, because the BIARE variable refers to ``security'', rather than ``insecurity'' as in the previous regression using the ENVIPE survey variable. We can conclude that SWB is truly multidimensional, which is logical when we consider that satisfaction with life depends on many factors including health (refer to Subsection \ref{subs:basicdevelop}), family and community (refer to Section \ref{subs:individualism}) and many other factors (refer to Section \ref{subs:multidim}). 
%--------------------------------------------------------
\subsection{Gaussian graphical model for partial correlations}\label{subsec:ggmmicrodata}
%--------------------------------------------------------
Underlying a regression there is a particular structure of an explained variable and a set of regressors. In  estimations the coefficients involve correlations. This structure already suggests causality and gives a special role to a unique variable. In this section we change this paradigm and do not single out a variable but instead, analyze how a set of variables (selected questions from BIARE) are interconnected, inspired by the multidimensional nature of well-being. For these relationships we use partial correlations, and this is the key difference with a regression analysis. A partial correlation other than zero between two variables counts as a connection. In the context of Gaussian graphical models which we are going to consider here, such a connection is an edge in a graph constructed from the matrix of partial correlations. Shifting from correlations to partial correlations has the advantage that a connection between two questions, cannot be due to the influence of a third factor (question). Indeed, as we are going to show, this point of view will result in a new insight in the complex construct well-being. From an econometric point of view, we compute partial correlations through a Gaussian graphical model (refer to \cite{Borsboom2013}, for a survey on applications in psychology and to \cite{Lauritzen1996}, \cite{Maathuis2019}, for a systematic presentation of Gaussian graphical models). In this way we get direct relationships between questions which are not due to a third factor (i.e. through another concept in a third question). \\

\begin{figure}[H]
\begin{center}
\includegraphics[width=10cm,height=12cm]{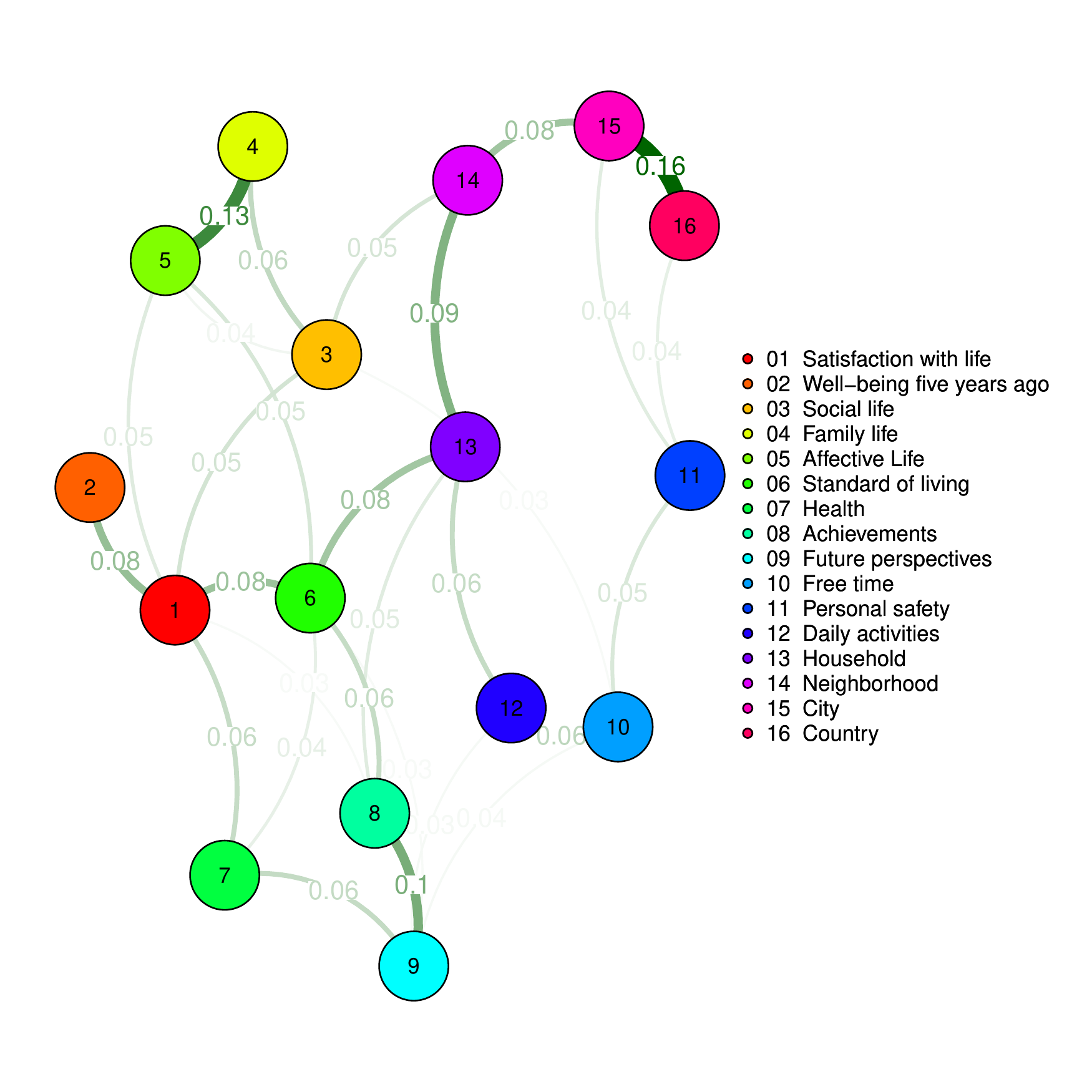}
\end{center}
\caption{Graph of Partial Correlations of Mexican Life Satisfaction According to BIARE data.}
\label{figGraphmodel}
\end{figure}

We consider two blocks of questions. The first block consists of the questions listed  in the Table \ref{table:biare1}. In Figure \ref{figGraphmodel} we see a graphic representation of the results in which nodes represent the questions and the width of the connection represents the magnitude of a partial correlation between corresponding nodes. The partial correlations are also written as labels. In particular, if two nodes are not linked through an edge it means that their partial correlation has a value between zero and the (somewhat arbitrarily low) threshold 0.03 and that they are at best indirectly connected through another node.
Interpreting the graph, we start with the geographical context, where we find an interesting path with nodes 16, 15, 14 and 13 representing satisfaction with the country, city, neighborhood and household, respectively, which may provide evidence of a community aspect to well-being (Refer to Section \ref{subs:individualism}). The strongest partial correlation in the graph is in fact between nodes 15 (city) and 16 (country). Another interesting component is given by nodes 7, 8 and 9, in which future perspectives (node 9) is related to achievements in life (node 8) but also to (good) health (node 7) (Refer to Section \ref{subs:basicdevelop} for the importance of health). Nodes 4 and 5 represent satisfaction with family life and satisfaction with affective life, respectively, which are of course naturally related (Refer to Section \ref{subs:individualism}). Surprisingly family does not link directly with satisfaction with life.  In another component of the graph, node 1 is directly connected to nodes 2, 3, 5, 6, and 7. Hence, satisfaction with life is directly connected to memories of well-being from the past (five years previously), but also social and affective life, standard of living, and health. This demonstrates that satisfaction with life is \textit{directly} influenced by some factors including subjective emotions as well as a self-perception of success grounded in material well-being. However, not as much as with correlations which also account for indirect relationships. Partial-correlation links then ``construct'' road maps for indirect connections, for example, to the geographical context. There are also some surprises in contrast with established facts in the literature. For example, we would expect family and personal security to be connected to satisfaction with life, but such connections are not manifest as direct links.  Interestingly, personal security is poorly connected with other questions, which coincides with its uniformly poor correlations in the Correlogram Figure \ref{FigCorrelogram}.  \\

\begin{figure}[H]
\begin{center}
\includegraphics[width=10cm,height=12cm]{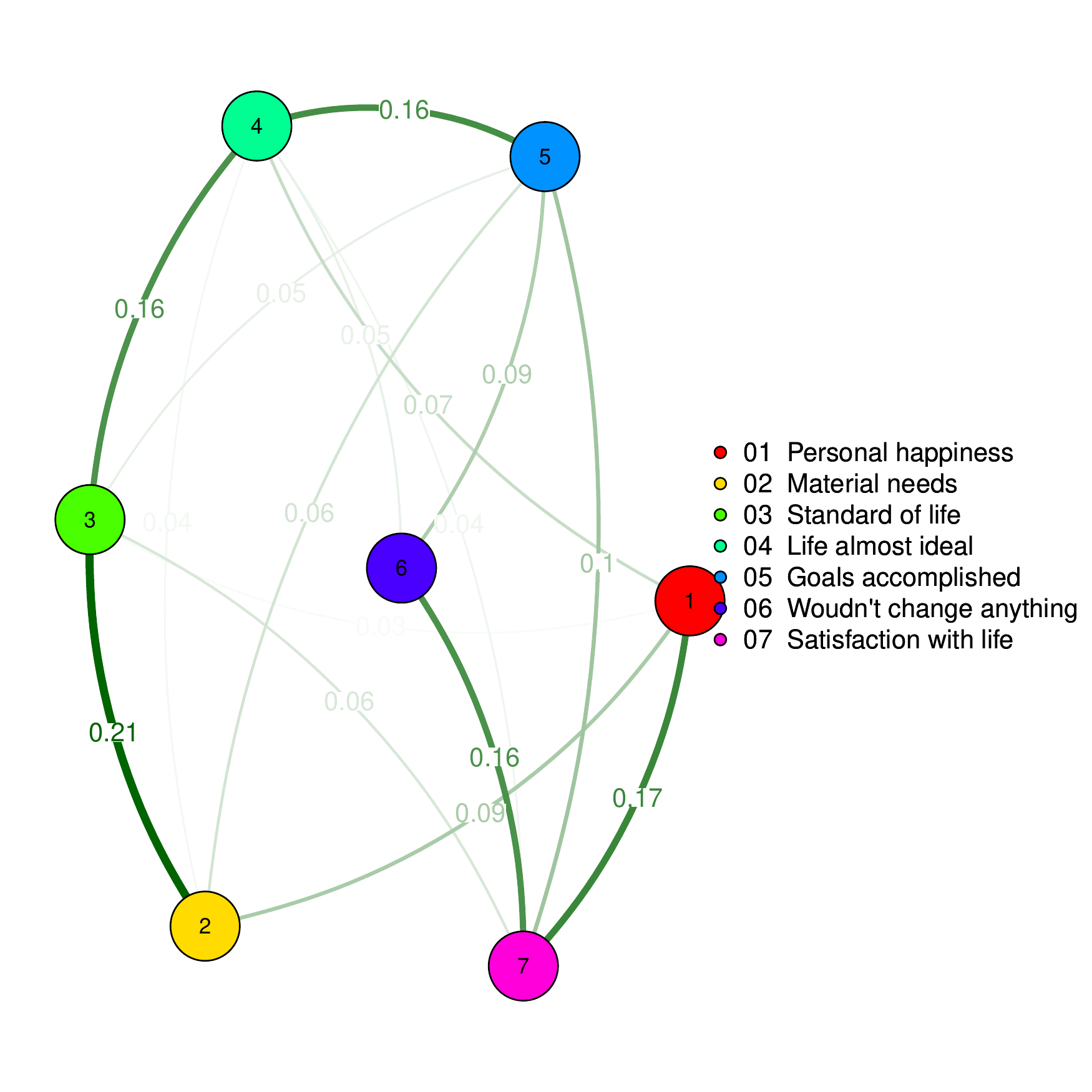}
\end{center}
\caption{Graph of Partial Correlations of Mexican Life Satisfaction According to BIARE data.}
\label{figGraphmodel2}
\end{figure}

For the second block we consider the questions listed in Table \ref{table:biare2}.  The graph of partial correlations is presented in Figure \ref{figGraphmodel2}. \\
Direct links with node 1, personal happiness, include nodes 2, 4 and 7, concerning material needs, life almost ideal and satisfaction with life, respectively. The strongest link is between node 1 and 7 and this is interesting since it connects hedonic SWB with evaluative SWB,  although being satisfied with life and happy at the same time is hardly surprising. The link between nodes 1 and 2 is weaker and again associates a hedonic experience with objective well-being and material needs. It would appear that material needs and standard of life do not guarantee happiness, and visa-versa, which concurs with Latin American happiness in adversity (refer to Section \ref{subs:paradox}). Another strong link is between nodes 6 ``wouldn't change anything'' and 7 ``satisfaction with life'', as one would expect. The link between nodes 4 and 5, ``life almost ideal'' and ``goals achieved'', respectively, is natural, and although in the model there is no in-built causality, one is tempted to suggest it here as an almost ideal life is surely related to meeting ones goals.. Nodes 2 and 3 are also naturally connected because satisfying material needs requires running a household, which improves the members' standard of life.
%--------------------------------------------------------
\subsection{Discussion}
%--------------------------------------------------------
Both SWB and OWB are now recognized as important for studies of happiness, quality of life, etc. (\cite{Rojas2011}, \cite{Fernandez2019}). While objective indicators have the advantage of being verifiable, they are indirect measures, they may be motivated by prevailing doctrine and not just science, and as such may not even be functional proxies, whereas subjective measures are direct, although their instant reply spontaneity, may make them unreliable, and they may suffer response heterogeneity due to cultural factors (\cite{Rojas2004}). In other words, both types of indicators have their drawbacks, so the sensible solution is to use both. Combined, as the Fitoussi report proposes, they can logically give more insight into the human condition than studies which exclude one or the other approaches (refer to Section \ref{subs:SSFreport}). As discussed in Section \ref{subs:descripAnl}, studies of well-being in Mexico and its federal entities (states), reveal strong geographical heterogeneity in terms of comparative development with southwestern states far poorer than their central and northern equivalents, which gives data sets the opportunity to explore how underdevelopment and poverty affects well-being (\cite{Fernandez2019}, \cite{Heald2018}). Data are also heterogeneous because those inequalities are mirrored at municipal level, creating an enormous gap between rich and poor. As Section \ref{subs:materialliving} discusses, international studies have found that at lower levels of development, income is important for well-being and happiness, but that it is only one of numerous relevant factors (\cite{Keny2005}). Satisfying material needs is also important, but as with income, there is evidence that there are diminishing returns (\cite{Cummins2018}, \cite{Eckersley2009}). Human beings are complex and relative income is important due to tendencies to compare with the neighbors (\cite{Camfield2010}). The literature also testifies that the multifaceted domain of health contributes to happiness and well-being (\cite{Cazzuffi2018}  \cite{Maccagnan2017}), which would appear entirely logical, although as Section  \ref{subs:basicdevelop} discusses, inexplicably it does not always appear to correlate well with other well-being domains (\cite{Heald2018}). As Section \ref{subs:basicdevelop} also discusses, the interpretation of education in the well-being literature is also complicated due to the aspirations of social climbers who may not reach their development goals (\cite{Clark2015}, \cite{Kristoffersen2018}). In other words, many studies portray education as QoL neutral, which is not corroborated by our analysis which reveals educational as intimately connected to well-being according to interstate data. There is some truth in the notion that Economic theory is biased towards the study of individual well-being rather than that of the community, due to its fixation with private property, perfect competition, etc., which means that the study of social capital is relatively recent, with important measurement problems, as discussed in Section  \ref{subs:individualism}. (\cite{Eckersley2013},  \cite{Hooghe2011}). The evidence of community well-being impacting on individuals via transmitters like inequality are poorly understood (\cite{Schneider2016}), which may explain why it appears to be QoL neutral in our analysis. Due to the multifaceted nature of well-being and happiness (refer to Section  \ref{subs:multidim}), many researchers build dimensions or domains of indicators, representing the diversity of influencers, including income, health, education, etc. (\cite{Rojas2006}, \cite{Rojas2018}). Mexican and Latin Americans reveal some interesting peculiarities in the literature, principally that they are happier than they should be (refer to Section  \ref{subs:paradox}), taking into account modest incomes and high crime rates and insecurity, apparently more than countered by strong extended family relationships which confer ample satisfaction and explains why Latins score high for happiness (\cite{Rojas2012}, \cite{Rojas2018}). Frustratingly, crime and insecurity indicators do not easily integrate into multidimensional well-being (refer to Section  \ref{sec:methodologies}).\\

Our descriptive research with Mexican data (refer to Section  \ref{subs:descripAnl}) confirms the observations of other Latin American studies discussed in Section \ref{sec:recentfindings}. Objective indicators are correlated over most domains as can be expected due to processes of cumulative causation in regional development, however in our analysis, many subjective indicators also correlate strongly with objective indicators, and subjective indicators correlate with each other, more so than in comparable studies (\cite{Fernandez2019}). We can interpret such concurrences as vindication for using subjective as well as objective indicators in QoL studies.  It means that what people say and perceive directly does have research validity and universality. On the other hand, inequality as measured by the GINI indicator appears to be a misnomer (refer to Figure \ref{FigCorrelogram}) which has no simple explanation, although inequality is a ubiquitous feature across Mexican society which may be one reason it does not correlate. Another reason is our inability to measure social or community well-being or simply human individualism and selfishness (refer to  Section \ref{subs:individualism} and \cite{Evans2019}). \\

Our econometric analysis also corroborates wider research findings. Using regression equations, we found significant relations in Mexican data between satisfying material needs and happiness, and between income and happiness i.e. between objective and subjective realms of well-being (\cite{Fernandez2019}). Income and material well-being proved to be more closely matched, which is to be expected because they are closer neighbors (refer to Figures \ref{figre00Full} and \ref{figre12Full}).\\ 

The perception of insecurity had the expected negative sign when put in a regression model alongside income to explain satisfaction with life, although not significant, which  reflects ambiguity in the data when analyzed both descriptively and econometrically (refer to  Figures \ref{FigCorrelogram}, \ref{FigCorrelogram}, \ref{figmylogit} and Section \ref{subs:paradox}). However, when the dependent variable (life satisfaction) is categorized into high and low, personal security does turn out to significantly impact on life satisfaction, alongside other SWB indicators. The reason, which is also observable in the bubble diagram in Figure \ref{FigBubbleplot}, is that poor households appear to perceive the problem of security more acutely than higher-income households (\cite{Charles2018}). By analyzing at household level, the problem of security manifests itself more in poor in disadvantaged neighborhoods in which QoL and happiness is compromised by a plethora of impediments to well-being. 
Well-being however is multidimensional as evidenced in a multiple regression of BIARE and  SWB indicators (refer to Figure \ref{figmylogit}), which is corroborated in the top, left quadrant of the Correlogram, Figure \ref{FigCorrelogram}.
Finally, an analysis of SWB indicators using partial correlations demonstrates that well-being is truly multidimensional, including subjective (happiness) and objective (material needs) well-being, which also link together affective life, family life, household and neighborhood factors, and past well-being and future perspectives (refer to Figures  \ref{figGraphmodel}  and \ref{figGraphmodel2}, and Section \ref{subs:multidim}). 

Absences of partial correlations demonstrates that many of the correlations between BIARE SWB indicators and census OWB indicators reported in Section \ref{subs:descripAnl} are actually due to third factors. Similarly, the node diagrams of partial correlations in Figures \ref{figGraphmodel} and \ref{figGraphmodel2} show both strong connections and weak or absent ones. Thus, a complex structure. For example, we might expect satisfaction with life and perhaps happiness to be hubs or interconnections. In Figure 6, affective life and family life appear almost divorced from the rest of the SWB indicators and in Figure \ref{figGraphmodel2}, the nodes separate into two weakly connected groups. 
%--------------------------------------------------------
\section{Conclusion}\label{sec:conclusion}
%--------------------------------------------------------
Subjective indicators are proving useful for the study of happiness, quality of life and well-being in general. They complement  objective measures and offers new avenues and opportunities for analysis. As the Fitoussi Report confirms, OWB is a powerful instrument but not the only one, suffering the drawbacks of being an indirect, imputed measure of well-being. SWB is direct and plain-spoken, although it suffers its demons too, including spontaneity (the transient mood of the interviewee), and cultural heterogeneity (some cultures may grade a Likert scale differently than others). An example of ideology creeping into analysis is perhaps a Western predilection for individual rather than community well-being, exposed in interviewing techniques which switch the interviewee away from community concerns to individual preoccupations. There are of course other sources of potential bias, for example, the selection of indicator domains or dimensions is normative, whether subjective or objective,  while the selection of mis-specified or incomplete analytical models will fail. We justify the use of exploratory correlation analysis, simple regressions, a logit model using categorized subjective indicator data, and partial correlations via Gaussian Graphical models to reveal some interesting results from our data. Subjective and objective measures of happiness and quality of life do correlate widely, so a multidimensional space for well-being clearly identifies itself. Middle-income Mexico, a deeply inequal geographical space with hardships provoked by economic and social realities, such as crime and insecurity, proves to be significantly happier and more satisfied than it should be according to its income,  as identified by the regional well-being literature and confirmed by our analysis. We find in our descriptive and quantitative analysis that relatively high Mexican household well-being is the consequence of a complex combination of factors linking subjective and objective features of well-being, in which income is required to provide for material needs, but in which household and standard of life, good health, past well-being and future perspectives may be as or more important in a multidimensional well-being framework. An intriguing aspect of our results is that there are also missing relationships i.e. indicators which are unexpectedly not directly related with the rest, for which there is no simple explanation and requires future research.

\appendix
\section{Blocks of Biare questions}\label{appendix:blocks}
% latex table generated in R 3.6.3 by xtable 1.8-4 package
% Mon Apr 20 00:46:13 2020
\begin{table}[H]
	\caption{First block of questions from BIARE.}\label{table:biare1}	
	\centering
	\begin{tabular}{ll}
		\hline
		Code & \mbox{Question graded 0-10 where 10 denotes maximum satisfaction} \\ 
		\hline
		encsat\_1 & How satisfied are you with your current life. \\ 
		encsat\_2 & How satisfied were you with your life five years ago? \\ 
		satis\_1 & How satisfied are you with your social life? \\ 
		satis\_2 & How satisfied are you with your family life? \\ 
		satis\_3 & How satisfied are you with your affective life? \\ 
		satis\_4 & How satisfied are you with your standard of living? \\ 
		satis\_5 & How satisfied are you with your health? \\ 
		satis\_6 & How satisfied are you with your life achievements? \\ 
		satis\_7 & How satisfied are you with your future perspectives? \\ 
		satis\_8 & How satisifed are you with the time you spend on what you like doing? \\ 
		satis\_9 & How satisfied are you with your personal security? \\ 
		satis\_10 & How satisfied are you with your daily activities? \\ 
		satis\_11 & How satisfied are you with your house? \\ 
		satis\_12 & How satisifed are you with your neighborhood? \\ 
		satis\_13 & How satisifed are you with your city? \\ 
		satis\_14 & How satisfied are you with your country? \\ 
		\hline
	\end{tabular}
\end{table}

\begin{table}[ht]
	\caption{Second block of questions from BIARE.}\label{table:biare2}	
	\centering
	\begin{tabular}{ll}
		\hline
		Code & \mbox{ Question graded 1-7 where  7 denotes in total agreement.} \\ 
		\hline
		afirma\_1 & You are a happy person. \\ 
		afirma\_2 & Your most important material needs are satisfied. \\ 
		afirma\_3 & Your standard of life is excellent. \\ 
		afirma\_4 & Your life is almost ideal. \\ 
		afirma\_5 & You have achieved the most important goals of your life. \\ 
		afirma\_6 & You wouldn´t change anything if you were born again. \\ 
		afirma\_7 & You are satisfied with your life. \\ 
		\hline
	\end{tabular}
\end{table}

\bibliographystyle{plain}
\bibliography{biblio}   % name your BibTeX data base

\end{document}